%% file: EXO-15-005_temp.tex
\pdfoutput=1

\documentclass[11pt,twoside,a4paper,cmspaper,final,collab]{cms-tdr}
\def\svnVersion{394707}\def\cmsCernNoTag{CERN-EP/2016-209}\def\cmsCernDate{\today}\def\cmsMessage{Published in Physics Letters B as \href{http://dx.doi.org/10.1016/j.physletb.2017.02.010}{\doi{10.1016/j.physletb.2017.02.010.}}}

\begin{document}\cmsNoteHeader{EXO-15-005}

\hyphenation{had-ron-i-za-tion}
\hyphenation{cal-or-i-me-ter}
\hyphenation{de-vices}
\RCS$Revision: 394686 $
\RCS$HeadURL: svn+ssh://svn.cern.ch/reps/tdr2/papers/EXO-15-005/trunk/EXO-15-005.tex $
\RCS$Id: EXO-15-005.tex 394686 2017-03-17 14:12:00Z alfloren $

\newcommand{\anaLumiee}{2.7}
\newcommand{\anaLumimumu}{2.9}
\newcommand{\SSMwidth}{80}
\newcommand{\Esixwidth}{14}
\newcommand{\limitZssm}{3.37}
\newcommand{\limitZpsi}{2.82}
\newcommand{\limitGone}{1.46}
\newcommand{\limitGthree}{3.11}
\newcommand{\cPZg}{\ensuremath{\cmsSymbolFace{Z}/\gamma^{*}}\xspace}
\newcommand{\GKK}{\ensuremath{\mathrm{G}_\mathrm{KK}}\xspace}
\newcommand{\ZPSSM}{\ensuremath{\cPZpr_\mathrm{SSM}}\xspace}
\newcommand{\ZPPSI}{\ensuremath{\cPZpr_\psi}\xspace}
\newlength\cmsFigWidth
\ifthenelse{\boolean{cms@external}}{\setlength\cmsFigWidth{0.7\textwidth}}{\setlength\cmsFigWidth{0.7\textwidth}}
\ifthenelse{\boolean{cms@external}}{\providecommand{\cmsLeft}{top\xspace}}{\providecommand{\cmsLeft}{left\xspace}}
\ifthenelse{\boolean{cms@external}}{\providecommand{\cmsRight}{bottom\xspace}}{\providecommand{\cmsRight}{right\xspace}}
\cmsNoteHeader{EXO-12-061}

\title{Search for narrow resonances in dilepton mass spectra in proton-proton collisions at $\sqrt{s}=13\TeV$ and combination with 8\TeV data }

\date{\today}

\abstract{

A search for narrow resonances in dielectron and dimuon invariant mass
spectra has been performed using data obtained from proton-proton
collisions at $\sqrt{s}=13\TeV$ collected with the CMS detector. The
integrated luminosity for the dielectron sample is \anaLumiee\fbinv
and for the dimuon sample \anaLumimumu\fbinv. The sensitivity of the
search is increased by combining these data with a previously analysed
set of data obtained at $\sqrt{s}=8\TeV$ and corresponding to a
luminosity of 20\fbinv. No evidence for non-standard-model physics is
found, either in the 13\TeV data set alone, or in the combined data
set. Upper limits on the product of production cross section and
branching fraction have also been calculated in a model-independent
manner to enable interpretation in models predicting a narrow
dielectron or dimuon resonance structure. Limits are set on the masses
of hypothetical particles that could appear in new-physics scenarios.
For the \ZPSSM particle, which arises in the sequential standard
model, and for the superstring inspired \ZPPSI particle, 95\%
confidence level lower mass limits for the combined data sets and
combined channels are found to be $\limitZssm$ and \limitZpsi\TeV,
respectively. The corresponding limits for the lightest Kaluza--Klein
graviton arising in the Randall--Sundrum model of extra dimensions
with coupling parameters 0.01 and 0.10 are $\limitGone$
and \limitGthree\TeV, respectively. These results significantly exceed
the limits based on the 8\TeV LHC data.  }

\hypersetup{%
pdfauthor={CMS Collaboration},%
pdftitle={Search for narrow resonances in dilepton mass spectra in proton-proton collisions at sqrt(s) = 13 TeV and combination with 8 TeV data},%
pdfsubject={CMS},%
pdfkeywords={CMS, dileptons, narrow resonances, extra dimensions}}

\maketitle

\section{Introduction}

The observation of a new narrow resonance in the invariant mass spectrum of lepton pairs would provide compelling evidence for physics beyond the standard model (SM). Many models designed to
address the shortcomings of the SM~\cite{Ellis:2012rsta,Kazakov:2014ufa} predict such
resonances at the \TeV scale. Examples include a new heavy $\cPZ$
boson-like particle such as the \ZPSSM boson of the sequential
standard model~\cite{Altar:1989}; the \ZPPSI boson inspired by
superstring models~\cite{Leike:1998wr,Zp_PSI_3}; and the Kaluza--Klein
graviton (\GKK) of the Randall--Sundrum (RS) model of extra
dimensions~\cite{Randall:1999vf, Randall:1999ee}.

This Letter describes a search for such narrow resonances in
dielectron and dimuon mass spectra based on proton-proton ($\Pp\Pp$)
collision data collected at $\sqrt{s}=13\TeV$ in 2015 by the CMS
experiment at the CERN LHC. The data correspond to integrated
luminosities of $\anaLumiee$ and $\anaLumimumu$\fbinv for the
dielectron and dimuon channels, respectively.  The ATLAS and CMS
Collaborations have previously reported searches in these
channels~\cite{Aad:2014cka,Khachatryan:2014fba} based on approximately
20\fbinv of $\Pp\Pp$ collisions at $\sqrt{s}=8\TeV$ in each
experiment. These results each exclude a \ZPSSM with a mass less than
2.90\TeV, and also exclude a \ZPPSI with a mass less than 2.51\TeV for
ATLAS and 2.57\TeV for CMS. Recently the ATLAS Collaboration has
increased these limits to 3.36\TeV ($\ZPSSM$) and 2.74 ($\ZPPSI$)
based on data from their first year of running at
13\TeV~\cite{Aaboud:2016cth}.

The data-taking and data-analysis methods for the 13\TeV data follow
closely those for the 8\TeV data~\cite{Khachatryan:2014fba}, with some
differences due to data-taking conditions and refinements noted below.
This Letter presents the search results from the 13\TeV data, followed
by results from combining the CMS data sets at 8 and 13\TeV; the
latter have only slightly more power, as most of the sensitivity at
high mass comes from the higher $\sqrt{s}$. As in previous searches,
the dimuon selection requires opposite sign charge for the muons,
while the dielectron selection has no sign requirement.

The primary results of the analysis are expressed in terms of the
ratio of the product of production cross section and branching fraction for a
possible new resonance to that for the $\cPZ$ boson. To determine this
ratio, the measured lepton pair invariant mass distributions are fit
to models that contain signal and background processes and incorporate
the ratio of efficiencies including the experimental acceptance. This
approach reduces the impact of many experimental and theoretical
systematic uncertainties. Furthermore, the analysis is designed to be
largely independent of specific model assumptions, enabling the
results to be interpreted in the context of any model that includes a
narrow spin-1 or spin-2 resonance decaying to an electron or muon
pair.  Here we present lower limits on the masses of hypothetical
particles that are derived from cross sections calculated in the
context of certain specific models.

\section{The CMS detector}\label{sec:CMS}

The central feature of the CMS detector is a superconducting solenoid
providing an axial magnetic field of 3.8\unit{T} and enclosing an
inner tracker, an electromagnetic calorimeter (ECAL), and a hadron
calorimeter (HCAL). The inner tracker is composed of a silicon pixel
detector and a silicon strip tracker, and measures charged-particle
trajectories in the pseudorapidity range $\abs{\eta}<2.5$.  The ECAL
and HCAL, each composed of a barrel and two endcap sections, extend
over the range $\abs{\eta} < 3.0$. The finely segmented ECAL consists
of nearly 76\,000 lead tungstate crystals while the HCAL is
constructed from alternating layers of brass and scintillator. Forward
hadron calorimeters encompass $3.0<\abs{\eta}<5.0$.  The muon detection
system covers $\abs{\eta}<2.4$ with up to four layers of
gas-ionization chambers installed outside the solenoid and sandwiched
between the layers of the steel flux-return yoke.  Additional
detectors and upgrades of electronics were installed before the
beginning of the 13\TeV data collection period in 2015, yielding
improved reconstruction performance for muons relative to the 8\TeV
data collection period in 2012. A more detailed description of the CMS
detector, together with a definition of the coordinate system used and
the relevant kinematic variables, can be found in
Ref.~\cite{:2008zzk}.

The CMS experiment has a two-level trigger system. The level-1 (L1)
trigger~\cite{L1TDR}, composed of custom hardware processors, selects
events of interest using information from the calorimeters and muon
detectors and reduces the readout rate from the 40\unit{MHz}
bunch-crossing frequency to a maximum of 100\unit{kHz}.  The software
based high-level trigger (HLT)~\cite{HLTTDR} uses the full event
information, including that from the inner tracker, to reduce the
event rate to the 1\unit{kHz} that is recorded.
\section{Event selection}

\subsection{Triggers}

The event selection and reconstruction algorithms employed are refined
versions of those used for previous high-mass dilepton
searches~\cite{Khachatryan:2014fba}.  The transverse energy of a
localized ECAL energy deposit (``cluster") is defined as
$\ET=E\sin\theta$, with $\theta$ the polar angle relative to the beam
axis, where the cluster energy $E$ includes deposits consistent with
bremsstrahlung emission. The selection of electrons begins with the L1
trigger, where electron candidates are defined as ECAL clusters with
$\ET>25\GeV$. In the HLT, electron candidates are defined as ECAL
clusters with $\ET>33\GeV$ that are matched to a track reconstructed
in the inner tracker. To suppress hadrons misidentified as electrons
in the barrel (endcaps), the energy deposited in the HCAL in a cone of
radius $\Delta R = \sqrt{\smash[b]{(\Delta\eta)^2 + (\Delta\phi)^2}} = 0.14$
around the electron candidate must be less than 15 (10)\% of the ECAL
cluster energy, where $\phi$ is the azimuthal angle. In the HLT,
events with at least two electron candidates are selected.

Muon candidates are identified with the L1 trigger by requiring each track
segment reconstructed in the muon detectors to have transverse
momentum \pt above 16\GeV. In the HLT, muon candidates are defined by
fitting hits from track segments in the muon detectors with hits from
segments in the inner tracker, with a \pt threshold on the track that
depended on the instantaneous luminosity and reached as high as 50\GeV
for unprescaled triggers.
The HLT muon candidates must have a distance of closest approach to
the beam axis less than 0.1\unit{cm} in the plane perpendicular to that
axis.  In the HLT, events with at least one muon candidate with $\pt >
50 $ GeV are selected.
To allow the normalization of rates, $\cPZ$ boson events are
obtained via a prescaled trigger that is identical to the primary analysis
trigger except that the \pt requirement is lowered to 27\GeV.

Trigger efficiencies are defined relative to the full analysis
requirements described in Section~3.2, and are evaluated from data
using high mass dilepton or high-\pt  $\cPZ$ samples, free from background contributions. For electrons with $\ET
>45\GeV$, the trigger efficiency of an electron pair is 99.6\% for
events with both electrons in the ECAL barrel, and 99.2\% for events
with one electron in the ECAL barrel and the other in an ECAL endcap,
and is consistent with being independent of \ET. For muons with
$\pt>53\GeV$, the trigger efficiency of a muon pair is 99.4\% and is
uniform in muon \pt.

\subsection{Lepton reconstruction}

The recorded events are processed with the CMS event reconstruction
algorithms~\cite{MUO-10-004-PAS,CMS-ele-paper}.

Electron candidates are defined by associating tracks in the inner
detector with ECAL clusters. The energy of the electron candidate is given
by the energy of the associated cluster, which is adjusted through
calibration and regression
methods~\cite{Cacciari:2007fd,Khachatryan:2014fba,CMS-ele-paper}. The associated
tracks provide the angular information used to calculate the electron
four-momentum. Each electron candidate must have $\ET>35\GeV$ and either
$\abs{\eta_{C}}<1.44$ (barrel region) or $1.56<\abs{\eta_{C}}<2.50$ (endcap
region), where $\eta_{C}$ is the pseudorapidity of the cluster with
respect to the nominal centre of the CMS detector. The electron
reconstruction efficiency is around 93\%~\cite{CMS-ele-paper} for
electrons within the acceptance region of the analysis.
At least two electron candidates are required for a dielectron event, at least
one of which must lie in the barrel region in order to exclude endcap-endcap events, which are dominated by the multijet background.

Muon candidate track segments are reconstructed separately in the muon
detector and inner tracker. Hits from a muon detector track segment
and from a compatible track segment in the inner tracker are fitted
under a global muon track hypothesis that incorporates information
from the entire CMS detector. Dedicated
algorithms~\cite{MUO-10-004-PAS}, developed for high-\pt (of the order
of 1\TeV) muon reconstruction, are needed to ensure the quality of the
hits contributing to the fit, as well as the quality of the fit
itself.
Events are required to contain at least two muon candidates, each with $\pt
> 53\GeV$, slightly above the corresponding HLT requirement, and to
appear within $\abs{\eta}<2.4$. The muon reconstruction efficiency for
muons within this region is above 98\%.

\subsection{Lepton identification}

Electron candidates are required to satisfy dedicated high-\ET
selection criteria~\cite{Khachatryan:2014fba}. The energy deposited in
the HCAL in a cone of radius $\Delta R = 0.14$ around the direction of
the electron candidate must be less than 5\% of the energy of the
electron measured in the ECAL.

Muon candidates are required to satisfy standard CMS muon selection
criteria, with modifications for high-\pt muon
identification~\cite{Khachatryan:2014fba} that emphasize information
from the muon detectors in order to improve the muon \pt resolution above 200\GeV.  Each pair of muon
candidates is fitted to a common vertex, with a requirement that the
resulting value of the $\chi^2$ per degree of freedom be less
than 20. This selection is designed to have an efficiency close to
100\% and to reject pairs formed from  mismatched muons.  To suppress
background from cosmic ray muons that pass near the interaction point,
the three-dimensional angle between the two track momentum vectors is
required to be less than $\pi-0.02$.

Finally, we impose isolation requirements to suppress jets
misidentified as leptons, and leptons from hadron decays. Electrons
are considered to be isolated if the \pt sum of tracks within a cone
of radius $\Delta R = 0.3$ around the direction of the candidate is
less than 5\GeV and if the \ET sum of energy deposits within this same
cone
less than 3\% of the candidate's \ET value, once corrected for the
contributions expected from detector noise and additional interactions
in the event~\cite{Khachatryan:2014fba}.  The majority of the dilepton
events in the analysed data set contain between 7 and 12 additional
interactions.  Similarly, muons are considered to be isolated
if the \pt sum of tracks within a cone of radius $\Delta R = 0.3$
around the candidate direction is less than 10\% of the \pt of
the candidate. The sums exclude the lepton candidate under
consideration.

The electron candidates in a dielectron event are not required to have
opposite charges because the charge misidentification rate is
non-negligible for high-\pt electrons. In contrast, we require muon
candidates in a dimuon event to have opposite charge because in this
case a charge mismeasurement, while rare, implies a large \pt
mismeasurement. If there are more than two electron candidates
selected in the event, the two highest-\pt electrons are used to
construct the pair. This procedure is also used when constructing a
dimuon pair.

The efficiency to select signal events, accounting for the effects of
event reconstruction, lepton identification and, in the case of muons,
the effect of the trigger, is determined from Monte Carlo (MC)
simulations.  Details of the simulation are given in
Section~4. Methods relying primarily on data, such as the use of
control samples of high-\pt $\cPZ$ bosons decaying to
$\Pep\Pem$ and $\PGmp\PGmm$ pairs, are employed to
validate the simulation up to muon $\pt= 300\GeV$. The simulated and
measured efficiencies generally agree within about 1\%, for both
single electrons and muons. High mass dilepton or high-\pt $\cPZ$
samples, where background sources are subtracted using MC information,
are used to extend the validation up to $\pt \approx
1\TeV$. Differences between data and simulation up to around 5\%
(2.5\%) for single electrons (muons) are found for these large \et (\pt)
values. While the muon trigger efficiency is estimated from
simulation, the efficiency of the primary electron trigger at
high mass can be estimated from data because of the presence of simple
calorimeter-based triggers that are fully
efficient for high mass electron pairs.These simple calorimeter-based
triggers have high \et thresholds, which prevent their use for the
entire mass range.

The signal efficiency within the acceptance of the analysis is found to be
($75\pm 8$)\% and ($70 \pm 10$)\%, respectively, for a barrel-barrel and
barrel-endcap electron pair of $1\TeV$ mass. For a muon pair with a mass
of $1\TeV$, the corresponding efficiency is $91^{+1}_{-5}\% $. The
uncertainties in the efficiency values account for the statistical
precision and for the systematic uncertainty in the extrapolation of
the data-simulation differences to high \pt. The acceptances are
derived from simulation and rise with increasing
mass. In the dimuon channel the probability for a produced boson
with 400\GeV mass to decay within the detector acceptance is
close to 40\% while for a 3\TeV mass it is
greater than 90\%. The acceptance is slightly lower in the dielectron
channel since endcap-endcap events are not considered.

\subsection{Mass resolution and scale}

The shape of the signal distribution in the dilepton mass is described
by the convolution of a Breit--Wigner (BW) function, describing the
intrinsic signal shape, and a Gaussian distribution, describing the
experimental resolution. As discussed in Section 5, the analysis is
insensitive to interference and similar effects. Note that for a resonance mass of 2.5\TeV,
the intrinsic widths of the \ZPSSM and \ZPPSI resonances are
$\SSMwidth$ and \Esixwidth\GeV, respectively.  For this same mass
value, the intrinsic width of the $\GKK$ resonance is 0.35\GeV for a
coupling parameter
$k/\overline{M}_\mathrm{Pl}$~\cite{Leike:1998wr,Zp_PSI_3} equal to
0.01, and 35\GeV for a coupling parameter equal to 0.10, where $k$ is
the warp factor of 4-dimensional anti-de Sitter space and
$\overline{M}_\mathrm{Pl}$ is the reduced Planck scale. The resolution
is determined from simulation as a function of the generated dilepton
mass. The resulting resolution function is validated with data, using $\cPZ$ boson events for the dielectron sample and
cosmic ray events for the dimuon sample. The dielectron resolution
function is adjusted on the basis of this comparison to agree with the
measured result.
The experimental mass resolution, defined as the standard deviation of
the Gaussian function divided by its most probable value, is 1.4\% (1.8\%)
for barrel-barrel (barrel-endcap) dielectron pairs with a mass of 1\TeV.
The resolution for dimuon pairs with a mass of 1\TeV is 3.2\%.

The response of the detector to leptons may evolve as the dilepton
mass increases.  For electrons this could arise from a nonlinear
response of the readout electronics. However, with the current data
set there is no evidence for such an effect and the energy scale of
electrons above 500\GeV is validated at the 1--2\%
level~\cite{CMS-ele-paper}.  As the muon \pt increases, its measurement
becomes increasingly sensitive to the detector alignment. New methods
have been developed for the 2015 data to determine a potential bias
from this source.  The curvature distributions of positive and
negative muons in data are compared to those obtained in simulation
for different $\eta$ and $\phi$ ranges.  The effects of misalignment
not already included in simulation are modelled with additional
smearing applied to the dimuon mass resolution. This is particularly
important for muons with $\abs{\eta} > 0.9$, since their \pt measurement
in this region cannot be validated with cosmic rays. The resulting
resolution for a dimuon pair with mass 1\TeV is increased from 3.2\%
to 3.8\% in order to account for a potential misalignment in the muon
system.  Finally, for dimuon pairs, an additional 1\% uncertainty is
assigned in the position of the mass peak to account for other
possible sources
of scale bias such as detector movement due to magnet cycles.

\section{Background estimation}

The principal SM background arises from Drell--Yan (DY) production
($\cPZg$) of $\Pep\Pem$ and $\PGmp\PGmm$ pairs.  Additional sources of
background are top quark-antiquark (\ttbar), single top quark (tW),
diboson (WW, WZ, and ZZ), and DY $\tau^+\tau^-$ production, although
the relative contributions of these sources diminish with increasing
dilepton mass.  Events in which at least one electron candidate is a
misidentified jet contribute a small background in the mass region of
interest.  The multijet background is negligible in the dimuon channel
where it is found to be less than 0.2\% for masses above 200\GeV, as
for the previous 8\TeV analysis~\cite{Khachatryan:2014fba}. The
contribution of cosmic ray events is also negligible.  An additional
SM source of $\Pep\Pem$ and $\mu^+\mu^-$ pairs comes
from the photon-induced process $\gamma\gamma \to
\ell^+\ell^-$~\cite{Bourilkov:2016qum}, where $\ell$ is an electron or
muon.  The theoretical predictions at\TeV mass scales for this process
have a significant uncertainty, with some predictions~\cite{Ball:2013hta} indicating that
the photon-induced process is the dominant source of dilepton pairs
with mass above 3 TeV. Even if the relative contribution of this process
to background at such high masses is large,
the absolute contribution is small and, as noted below, the potential
effect on the derived limits is negligible.

The background from DY, \ttbar, tW, and diboson events is evaluated
from simulation. Direct DY, \ttbar, and tW production are simulated
with the
\POWHEG~v2~\cite{Nason:2004rx,Frixione:2007vw,Alioli:2010xd,Alioli:2008gx,Frixione:2007nw,Re:2010bp}
next-to-leading order (NLO) event generator, with parton showering and
hadronization described by \PYTHIA~8.2~\cite{Sjostrand:2014zea}.
Diboson processes are simulated at leading order (LO) with
\PYTHIA, and DY $\tau^+\tau^-$ production at NLO with
\MADGRAPH5\_aMC@NLO 2.2.2~\cite{MadGraph5} interfaced with
\PYTHIA.
The NNPDF2.3LO~\cite{Ball:2012cx} parton distribution functions (PDFs)
are used for the diboson samples and the
NNPDF3.0NLO~\cite{Ball:2014uwa} PDFs are used for the rest of the
samples.
The PDFs are evaluated using the LHAPDF
library~\cite{Whalley:2005nh,Bourilkov:2006cj,Buckley:2014ana}.  The
detector response is simulated with the
\GEANTfour~\cite{Agostinelli:2002hh} package.

Over the full DY spectra multiplicative corrections are computed with
\FEWZ 3.1~\cite{Li:2012wna} to take into account missing contributions
like QCD effects at next-to-next-to-leading order (NNLO), electroweak
effects at NLO in addition to pure QED effects, and photon-induced
lepton pair production.  These corrections have a negligible impact on
the final results. The data and MC backgrounds are normalized to the
event yield in the Z boson peak region, so that the resulting
normalization is independent of the detector luminosity
calibration. For \ttbar, tW, diboson, and DY $\tau^+\tau^-$
production, the produced number of e$\mu$ final states should be equal to the
sum of $\Pe\Pe$ and $\PGm\PGm$ final states. This feature is used to compare
the $\Pe\mu$ spectrum with suitably scaled MC predictions. The resulting
scale factors are all consistent with unity and are not applied in the
analysis.

The background from jets misidentified as electrons is evaluated from
multijet data control samples. The method is the same as that
described in Ref.~\cite{Khachatryan:2014fba}, except that data
sidebands, rather than MC predictions, are used to evaluate the
contributions to the control samples from genuine electrons and
photons misidentified as electrons.  The method takes into account the
different ways in which one or two misidentified jets, in possible
conjunction with other particles, can satisfy the selection criteria
for dielectron events.

\section{Statistical analysis and results}

\begin{figure}[htbp]
\centering
\includegraphics[width=0.49\textwidth]{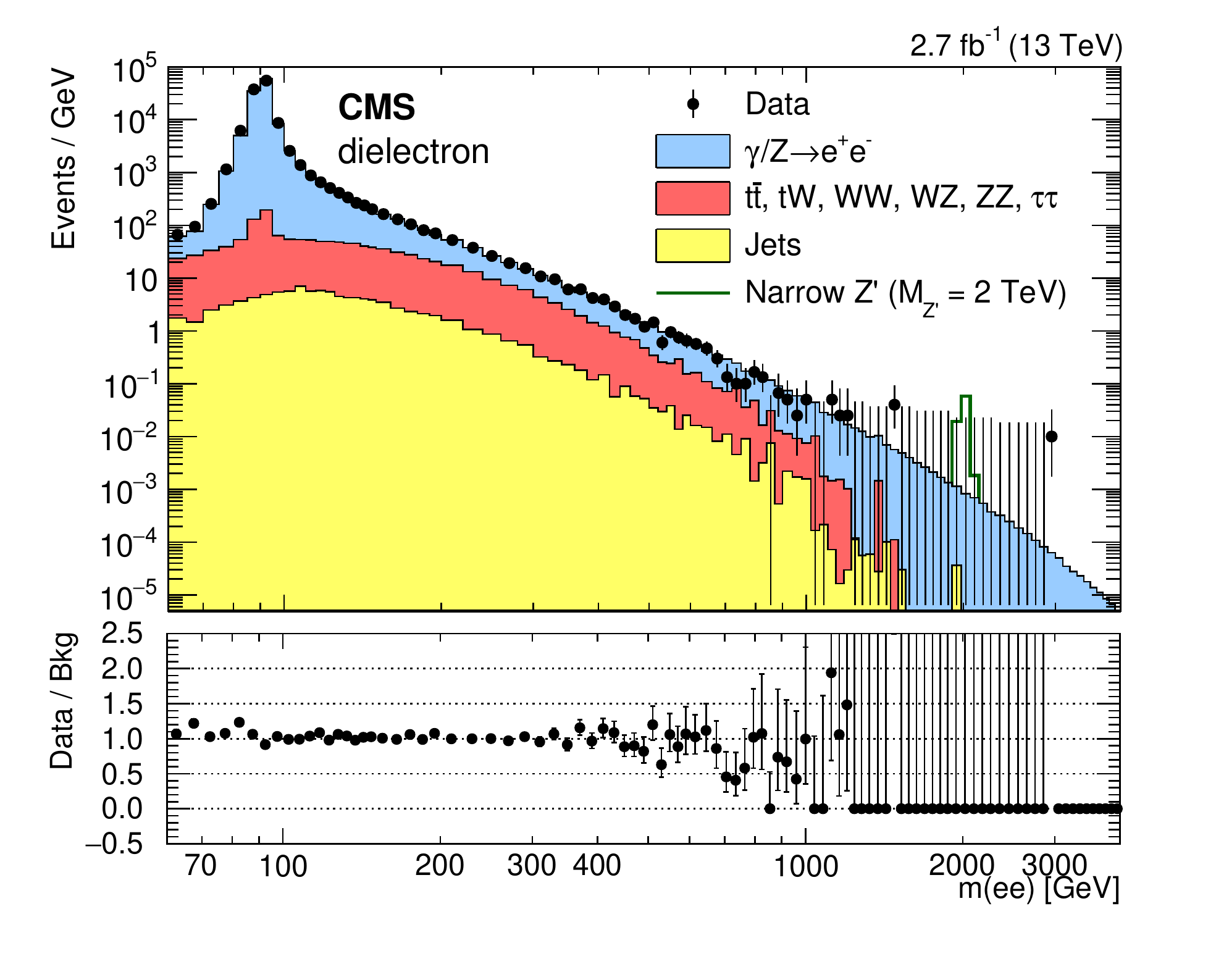}
\includegraphics[width=0.49\textwidth]{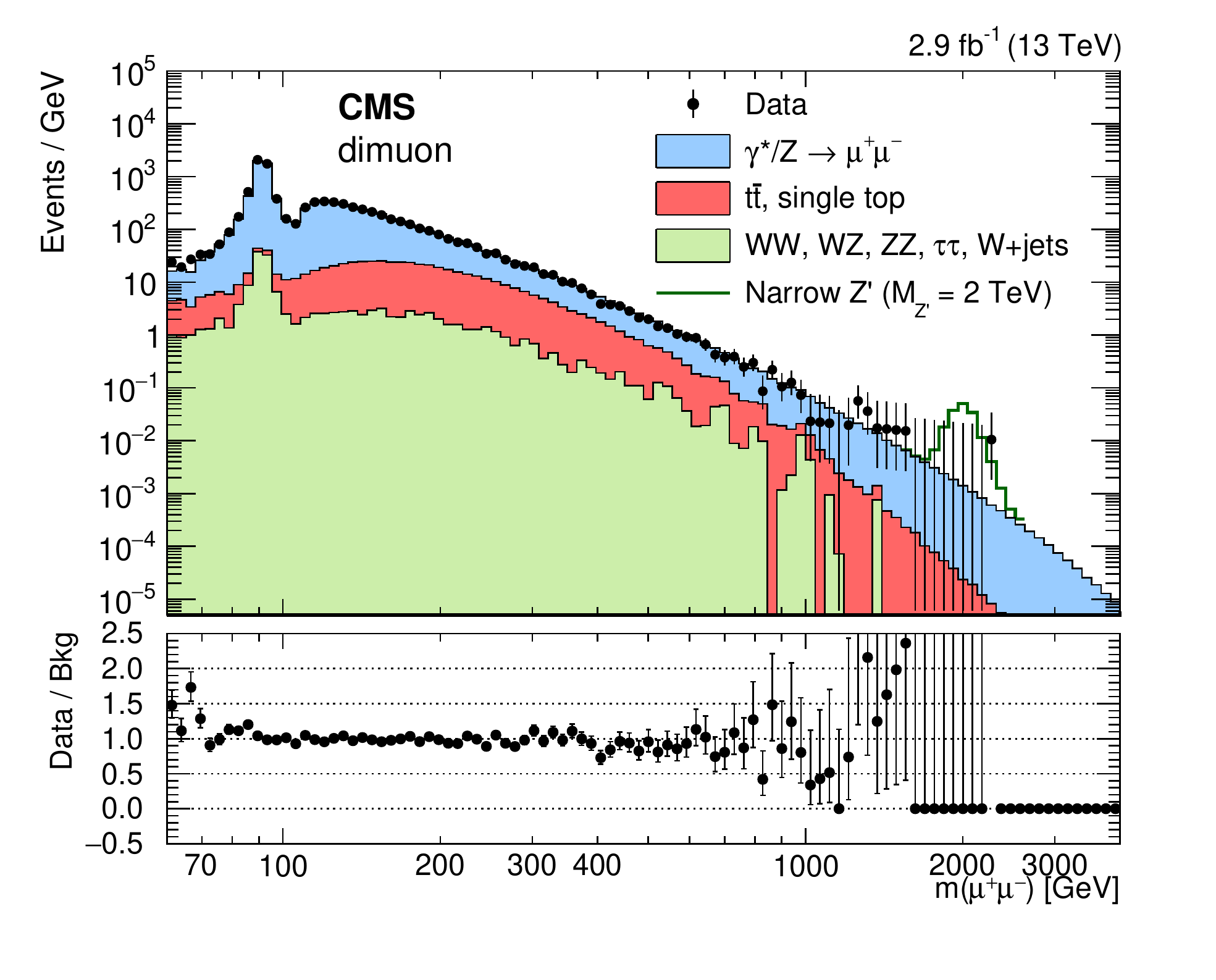}

\caption{ The invariant mass spectrum of (\cmsLeft) dielectron and
  (\cmsRight) dimuon events at $\sqrt{s}=13\TeV$.  The points with
  error bars represent the data.  The histograms represent the
  expectations from SM processes. The bins have equal width in
  logarithmic scale but the width in \GeV becomes larger with
  increasing mass. Example signal shapes for a narrow
  resonance with a mass of 2\TeV are shown by the stacked open
  histograms.  }
\label{fig:massSpectra}
\end{figure}

The observed invariant mass spectra of the dielectron and dimuon
events are presented in Fig.~\ref{fig:massSpectra}. No evidence for a
significant deviation from the SM expectations is observed. The
highest mass event observed is in the electron channel and has a mass
of 2.9\TeV. The estimated probability of observing a background event
with a mass at least this large is a few per cent in each channel.

Using a Bayesian approach with an unbinned extended likelihood
function~\cite{Khachatryan:2014fba}, limits are derived for the
production of a narrow spin-1 or spin-2 heavy resonance.  The
likelihood function is based on probability density functions (pdf)
that describe the signal and background contributions to the invariant
mass spectra. The signal distribution is parametrized by the
convolution of BW and Gaussian functions discussed in
Section~3.4. This analysis is designed for scenarios in which the BW
intrinsic width $\Gamma$ is small compared to the detector
resolution, and variations in $\Gamma$ therefore typically have little
effect on the derived limits. At high masses, however, the dielectron
mass resolution is comparable with the intrinsic width of the $\PZpr$ in
some of the models described in Section 3.4, and the limits can
exhibit some dependence on the assumed width.  Therefore results are
presented for different choices of the signal intrinsic width: 0.0,
0.6, and 3.0\% of the resonance mass.

The functional form of the background pdf is given by $m^{\kappa}\re^{\alpha m + \beta m^{2} +  \delta m^{3} }$ and is chosen to describe the
complete background representation produced using SM MC generators and
the background arising from misidentified jets deduced from the data.  For
each channel, the parameters of the background pdfs are obtained by
fitting the background distribution for masses above 400\GeV.

The limits are set on the parameter $R_{\sigma}$, which is the ratio
of the cross section for dilepton production through a $\PZpr$ boson
to the cross section for dilepton production through a \PZ boson:
\begin{equation}
\label{eq:rsigma}
R_\sigma = \frac{\sigma(\Pp\Pp\to \cPZpr+X\to\ell\ell+X)}
                {\sigma(\Pp\Pp\to \cPZ+X  \to\ell\ell+X)}.
\end{equation}

The Poisson mean of the signal yield is $\mu_\mathrm{S} = R_\sigma
\mu_\Z R_\epsilon$, where $R_\epsilon$ is the ratio of the selection
efficiency times detector acceptance for the $\zp$ decay relative to
that for the $\cPZ$ boson decay, and $\mu_\Z$ is the Poisson mean of
the number of $\cPZ\to\ell\ell$ events.  The value of $\mu_\Z$ is
estimated from the number of dilepton pairs in a $\pm$30\GeV window
around the $\cPZ$ boson mass, where the contributions of other
processes are predicted to be small (${\approx}0.5\%$ in
simulation). The quantities $\mu_\Z$ and $R_\epsilon$ are obtained
separately for the dimuon and dielectron channels.  By performing a
measurement relative to the Z boson cross section, the uncertainty in
the integrated luminosity is removed and uncertainties in other
quantities, such as in the experimental acceptance, trigger, and
reconstruction efficiencies, become relative rather than absolute.

The Bayesian limit-setting procedure follows closely that described
for the 8\TeV data in Ref.~\cite{Khachatryan:2014fba}.  The prior pdf
for the signal cross section is positive and uniform, as this is known
to result in good frequentist coverage properties.  Log-normal
functions are used to describe the systematic uncertainties.  Limits
on $R_\sigma$ are evaluated for scenarios in which the hypothetical
particle is either a spin-1 or a spin-2 resonance. The limits are sensitive to the number of
signal events relative to the number of background events, and to some
extent to the signal widths. Three classes of dilepton events are used
to set the limits: both electrons in the barrel section of the ECAL,
one electron in the barrel and the other in the endcap, and
dimuons. Dielectron events with an electron in the ECAL endcap are
studied separately because of their significantly higher multijet
background. To obtain the limit for a dilepton mass point, the amplitude of the
background shape function is constrained using data within a mass
window $\pm$6 times the mass resolution about the mass point.  If
fewer than 100 events in the 13\TeV data lie within this window
(rather than 400 used in the 8\TeV data), the window is symmetrically
expanded until this number is reached. This procedure sets the level
of the statistical uncertainty in the local background amplitude, and
the level is chosen to dominate expected systematic uncertainties in
the background shape at high mass. The uncertainties are larger in the
13\TeV data because of the reduction in the number of calibration
events due to the lower integrated luminosity, and because of the
higher mass ranges probed.  The observed limits are robust and do not
significantly change for reasonable variations in the limit-setting
procedure, such as modifications of the mass intervals used in the fit
or changes in the assumed background shape.

The limits obtained correspond to on-shell cross sections and do not
include model-dependent interference effects or enhancements at low
mass values related to the PDFs.  The limits are sensitive to the
fraction of events in each of three channels and so only apply to
models that contain a particle with the same spin as the particle in
the reference model, produced via a similar production mechanism. The limits are also only applicable to resonances with
widths of the order of a few per cent of the resonance mass, with the
limits becoming less applicable as the width increases.  Within these
constraints, the limits are, to a good approximation, model
independent and can be interpreted in the context of models not
explicitly addressed in this Letter.  A recipe to convert the cross
sections obtained from MC event generators such as \PYTHIA, which
include off-shell effects, to the on-shell cross sections presented
here is provided in Ref.~\cite{Accomando:2013sfa}.

\subsection{Combination of 8 and \texorpdfstring{13\TeV}{13 TeV} data sets}

The 13\TeV data set is combined with the 2012 data set at
$\sqrt{s}=8\TeV$~\cite{Khachatryan:2014fba}, corresponding to
integrated luminosities of 19.7 and 20.6\fbinv for the
dielectron and dimuon channels, respectively.
For the combination, these luminosities must be
rescaled to match the equivalent 13\TeV luminosities.  This scaling
depends on the mass of the resonance, with the effective luminosity of
the 8\TeV data sample decreasing with increasing resonance mass.  The
scaling was determined by comparing $\cPZpr$ and $\GKK$ cross sections
calculated by \PYTHIA using the NNPDF2.3LO PDF set at $\sqrt{s}=8$ and
13\TeV.  This cross section ratio depends on the PDF set used and
different choices of PDF set can change the resulting limits by a few
per cent.  The scaling also depends on the production mechanism of the
new boson, and therefore the value used to combine the data sets depends on the properties of the particular model under consideration.

The dominant uncertainty in this analysis is in the parameter
$R_\epsilon$. Its uncertainty is 8\% for the dielectron barrel-barrel
channel, 10\% for the dielectron barrel-endcap channel, and
$^{+1}_{-5}$\% for the dimuon channel.
The background from misidentified jets in the electron analysis is a
small fraction of the total background; therefore, although the
uncertainty in this background is large, its impact on the limit
determination is negligible. The uncertainty in the background shape
(which arises from uncertainties in the PDFs, in the contributions of
the photon-induced processes, and in the NNLO corrections to the cross
sections) is, as noted above, dominated by the statistical uncertainty in the background amplitude estimate. Possible photon-induced
contributions are studied using the MRST2004QED and NNPDF PDFs, which
include photons, and are found to have a negligible effect on the
derived mass limits. The uncertainty due to the PDFs is assessed using
the PDF4LHC15 prescription~\cite{Butterworth:2015oua} and is found to
vary from 2\% to 7\% as the dilepton mass increases from 1 to
4\TeV. Varying the numbers of background events within their total
uncertainties is found to have a negligible impact on the derived
limits. Common systematic uncertainties are taken to be fully
correlated in the calculation of combined limits. A relative mass
scale calibration uncertainty of 1\% is included when extracting the
combined limits using the 8 and 13\TeV data.
The uncertainties in the electron and muon efficiencies at high \pt are
taken to be uncorrelated between 8 and 13\TeV data, as most of these
uncertainties have their origin in calibration measurements made with
different data sets (with some variation as well in reconstruction and
identification variables used).

\subsection{Limits}

\begin{figure}[htb]
\centering
\includegraphics[width=0.49\textwidth]{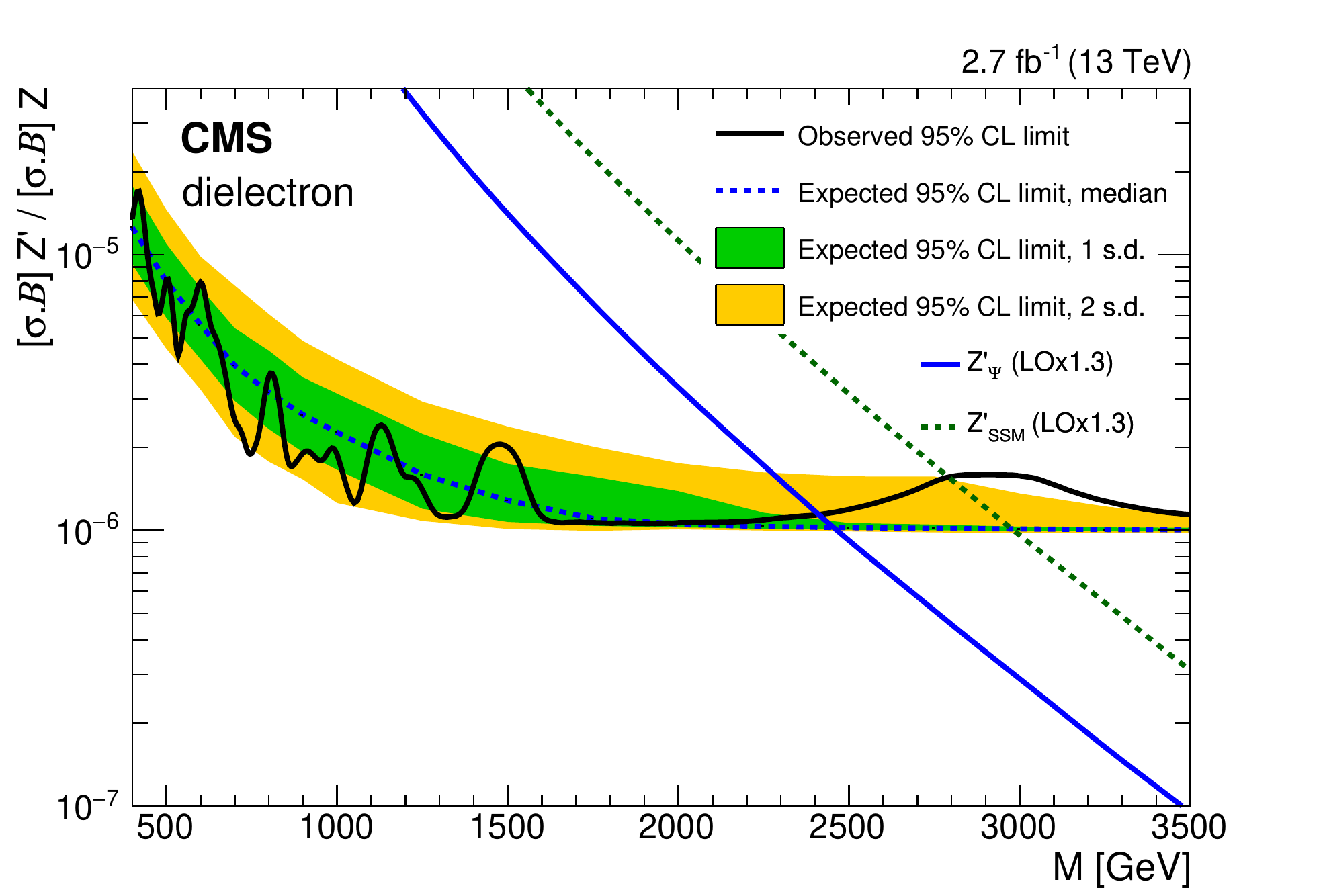}
\includegraphics[width=0.49\textwidth]{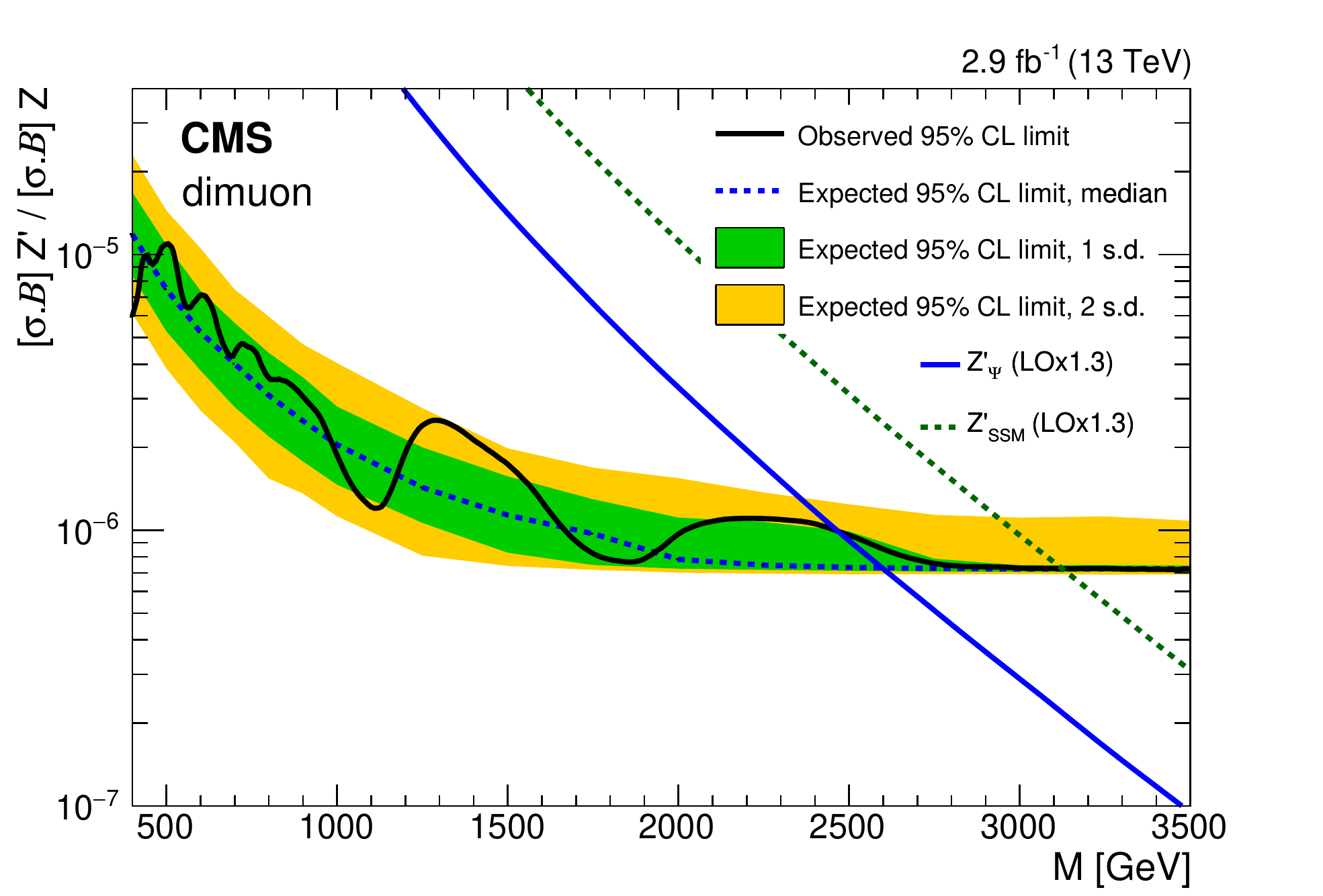}
 \caption{The 95\% CL upper limits on the product of production cross section
   and branching fraction for a spin-1 resonance with a width equal
   to 0.6\% of the resonance mass, relative to the product of production cross section
   and branching fraction for a Z boson, for the (\cmsLeft) dielectron
   and (\cmsRight) dimuon channels in the 13\TeV data. The shaded bands correspond to
   the 68 and 95\% quantiles for the expected limits.  Theoretical
   predictions for the spin-1 \ZPSSM and \ZPPSI resonances are
   shown for comparison.}
\label{fig:limits2}
\end{figure}

\begin{figure}[htb]
\centering
\includegraphics[width=0.49\textwidth]{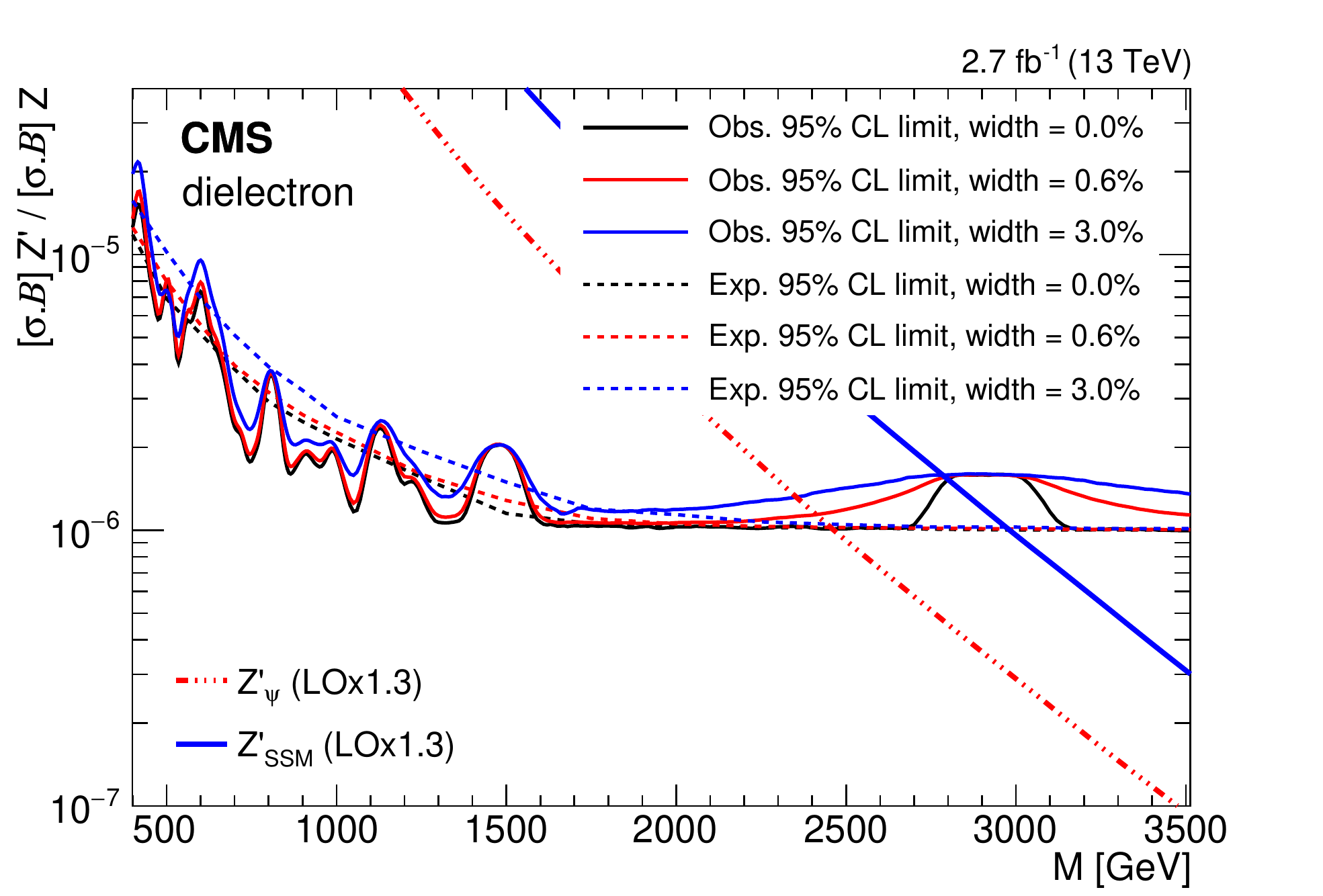}
\includegraphics[width=0.49\textwidth]{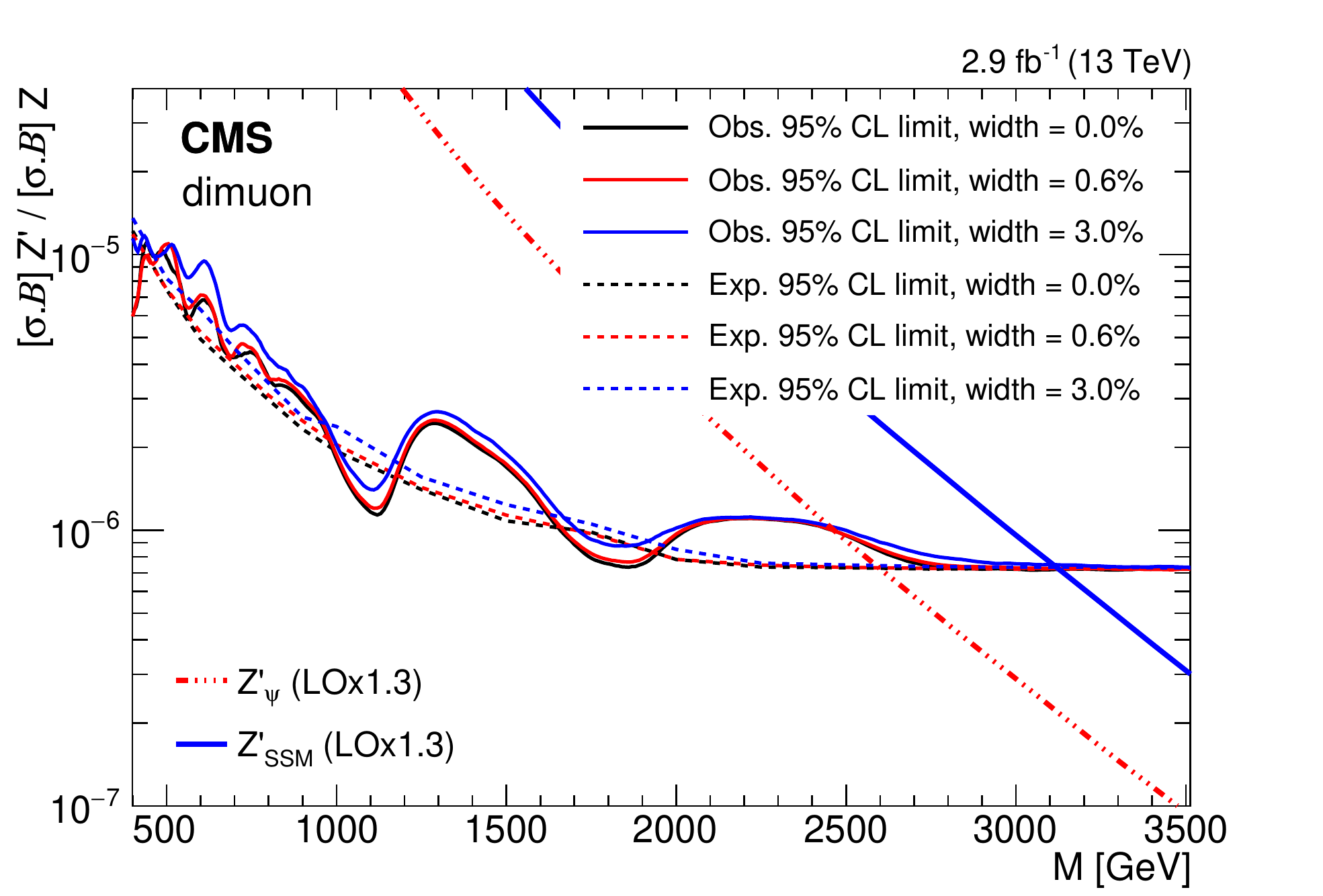}
 \caption{The 95\% CL upper limits on the product of production cross section
   and branching fraction for a spin-1 resonance for widths equal to
   0, 0.6, and 3.0\% of the resonance mass, relative to the product of
   production cross section and branching fraction for a Z boson, for the
   (\cmsLeft) dielectron and (\cmsRight) dimuon channels in the 13\TeV data.  Theoretical
   predictions for the spin-1 \ZPSSM and \ZPPSI resonances are
   also shown.}
\label{fig:limits3}
\end{figure}

\begin{figure}[htb]
\centering
\includegraphics[width=0.49\textwidth]{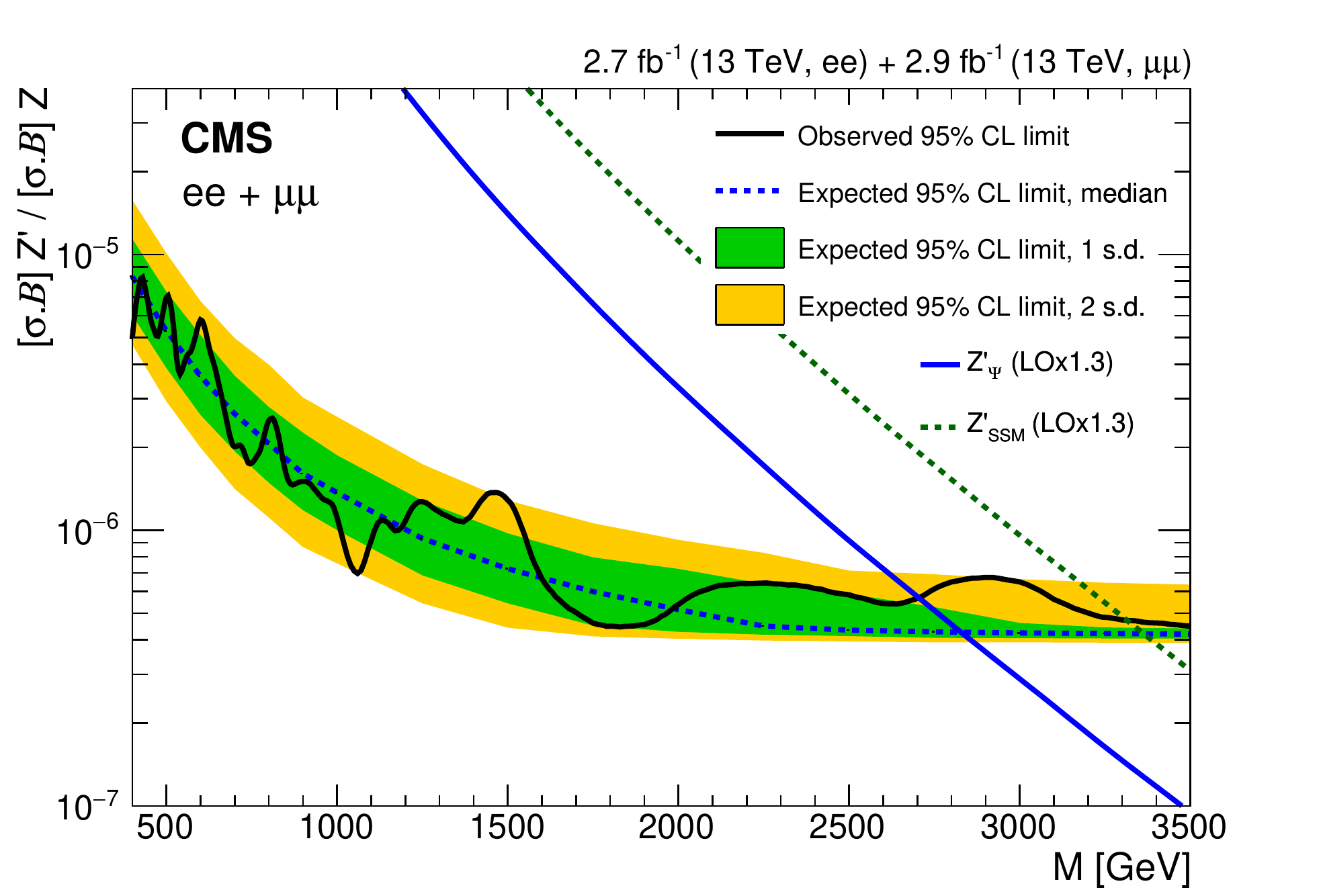}
\includegraphics[width=0.49\textwidth]{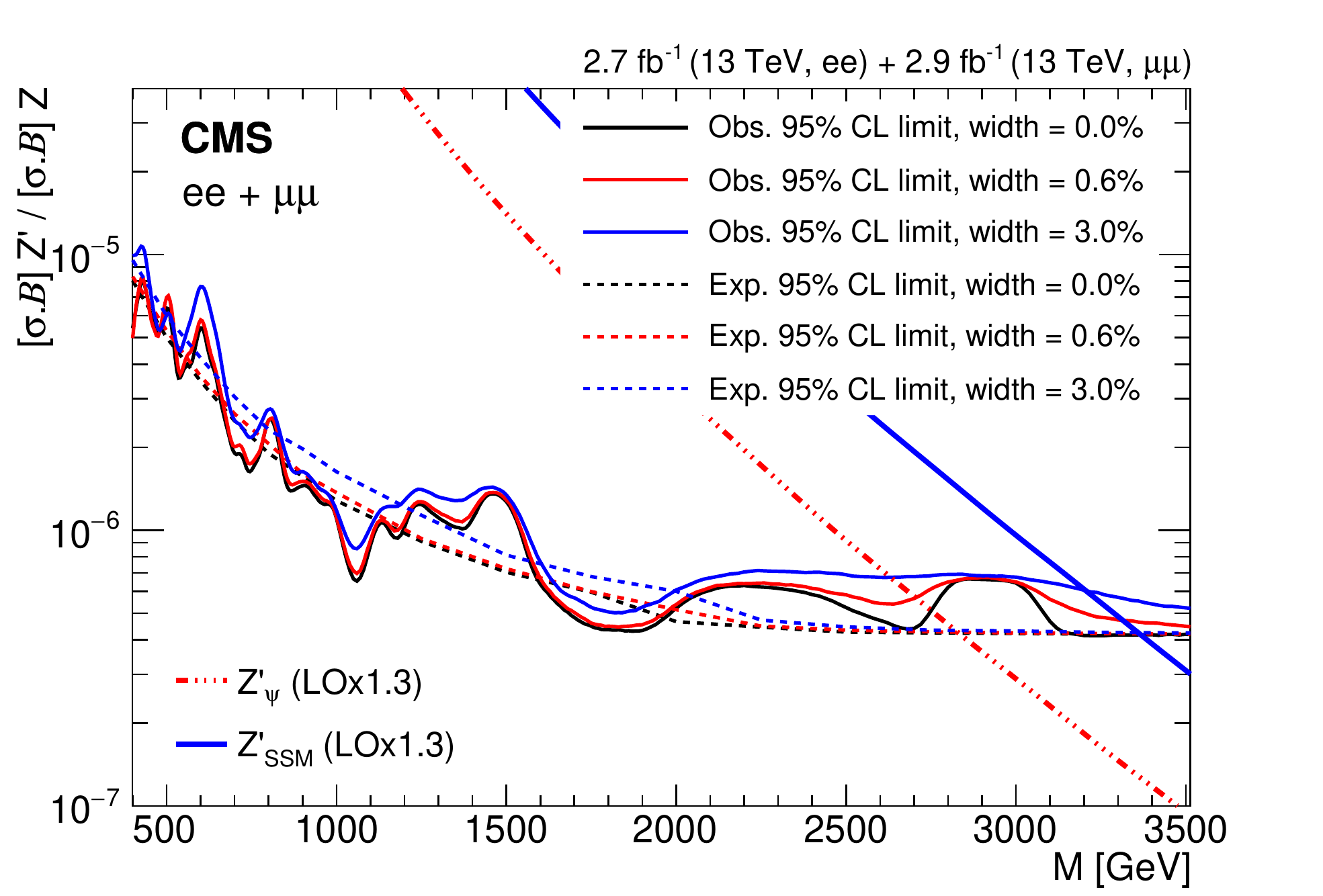}
 \caption{The 95\% CL upper limits on the product of production cross section
   and branching fraction for a spin-1 resonance, relative to the product of
   production cross section and branching fraction for a Z boson,
   for the combined dielectron and dimuon channels in the 13\TeV data,
   (\cmsLeft) for a resonance width equal to 0.6\% of the
   resonance mass and (\cmsRight) for resonance widths equal to 0, 0.6, and
   3.0\% of the resonance mass.  The shaded bands correspond
   to the 68 and 95\% quantiles for the expected limits.  Theoretical
   predictions for the spin-1 \ZPSSM and \ZPPSI resonances are
   also shown.  }
\label{fig:limits4}
\end{figure}

The 95\% confidence level (CL) upper limits on $R_{\sigma}$ for the 13\TeV
data are shown in Fig.~\ref{fig:limits2} for both the dielectron and
dimuon channels.
The resonance peak width for these results is taken to be 0.6\% of the
assumed mass value.  Results for widths equal to 0.0, 0.6, and 3\% of
the resonance mass are shown in
Fig.~\ref{fig:limits3}. Figure~\ref{fig:limits4} shows the 95\% CL
upper limits on $R_{\sigma}$ for the combination of the two channels
(assuming universality of electron and muon couplings) at 13\TeV
(\cmsLeft), and the corresponding effects of varying the signal width
(\cmsRight).

\begin{figure}[htb]
\centering
\includegraphics[width=0.49\textwidth]{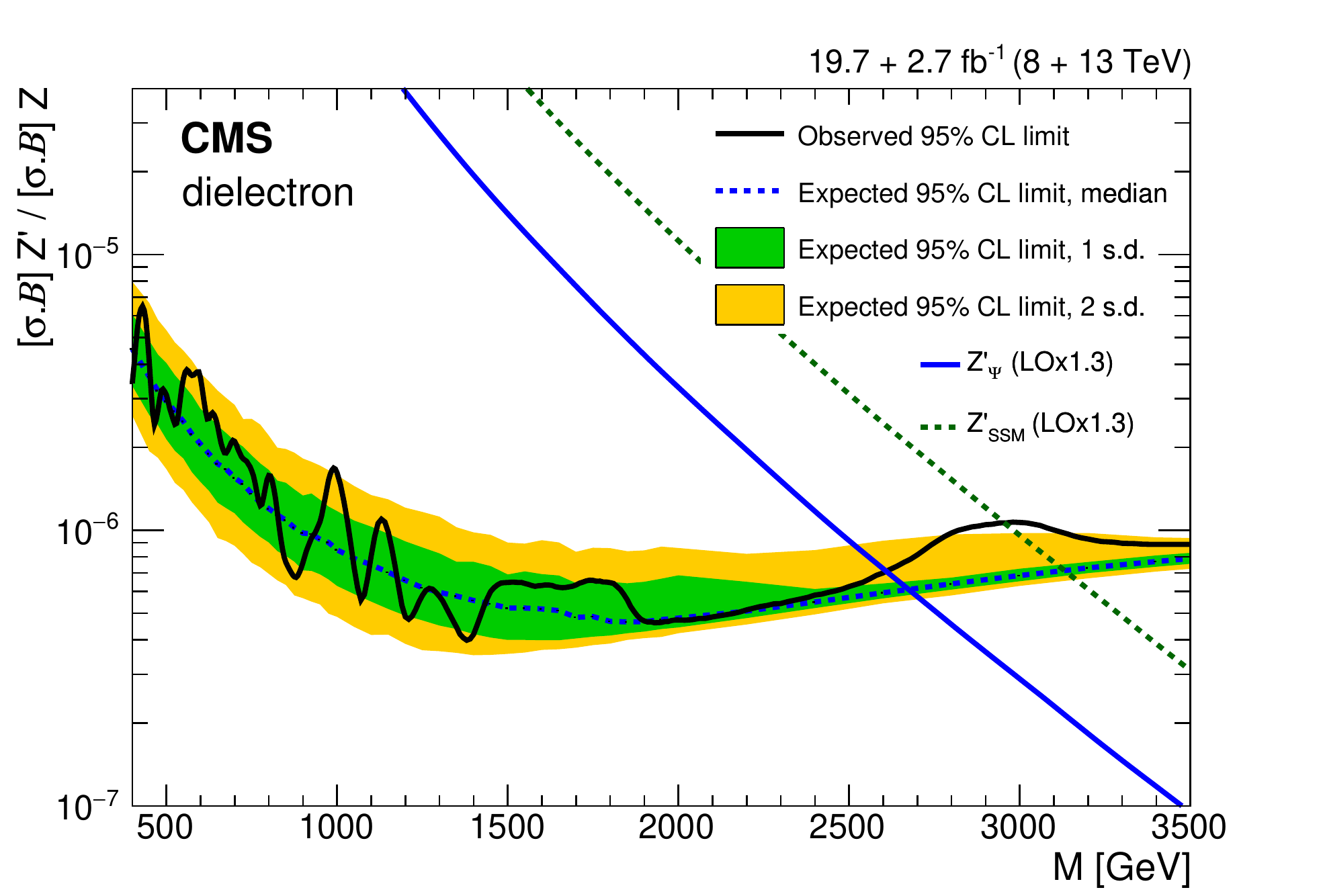}
\includegraphics[width=0.49\textwidth]{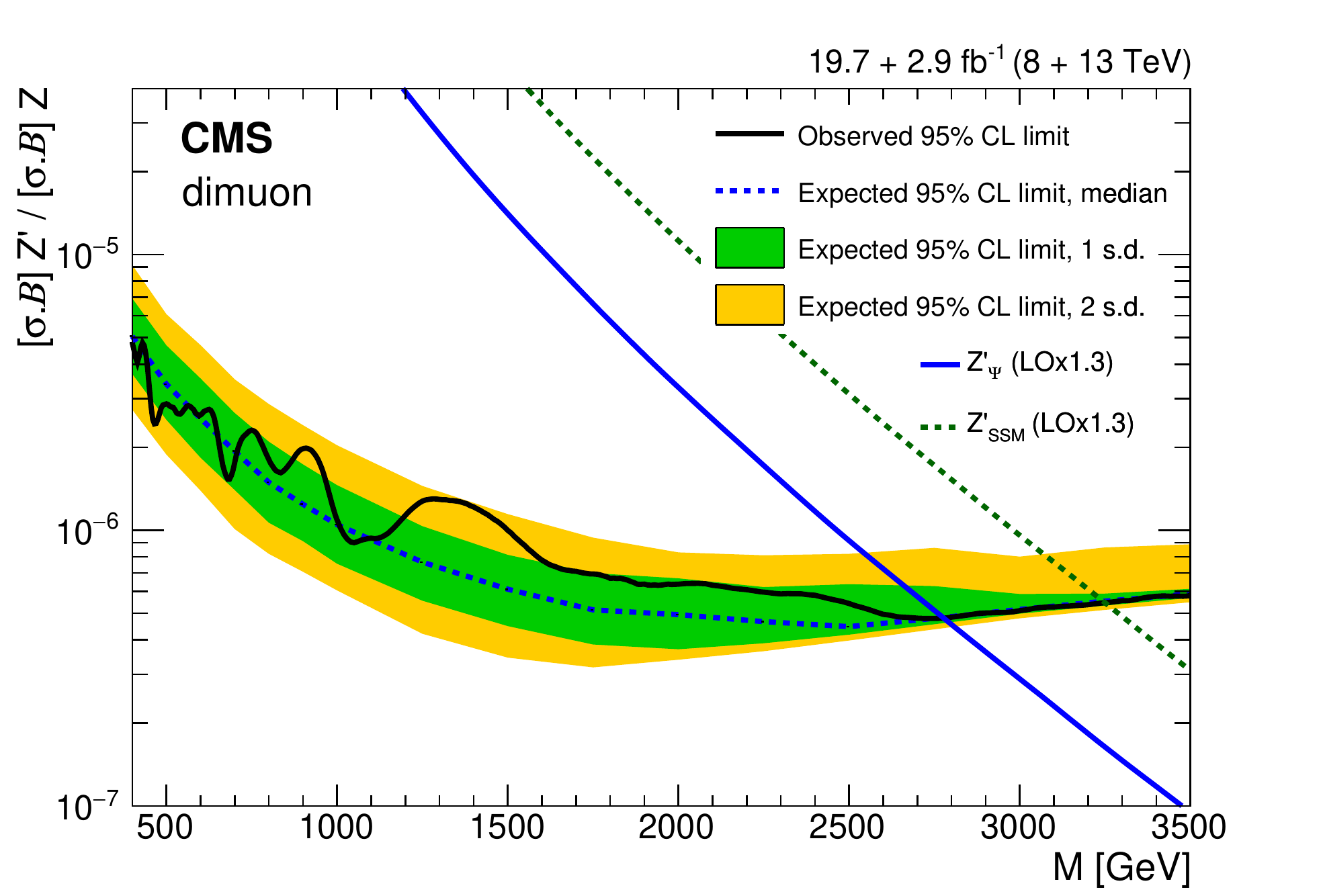}
 \caption{The 95\% CL upper limits on the product of production cross section
   and branching fraction for a spin-1 resonance with a width equal
   to 0.6\% of the resonance mass, relative to the product of production cross
   section and branching fraction for a Z boson, for the combined 8
   and 13\TeV data in the (\cmsLeft) dielectron and (\cmsRight) dimuon
   channel.  The shaded bands correspond to the 68 and 95\% quantiles
   for the expected limits.  Theoretical predictions for the spin-1
   \ZPSSM and \ZPPSI resonances are also shown.  }
 \label{fig:limits5}
\end{figure}

\begin{figure}[htb]
\centering
\includegraphics[width=0.49\textwidth]{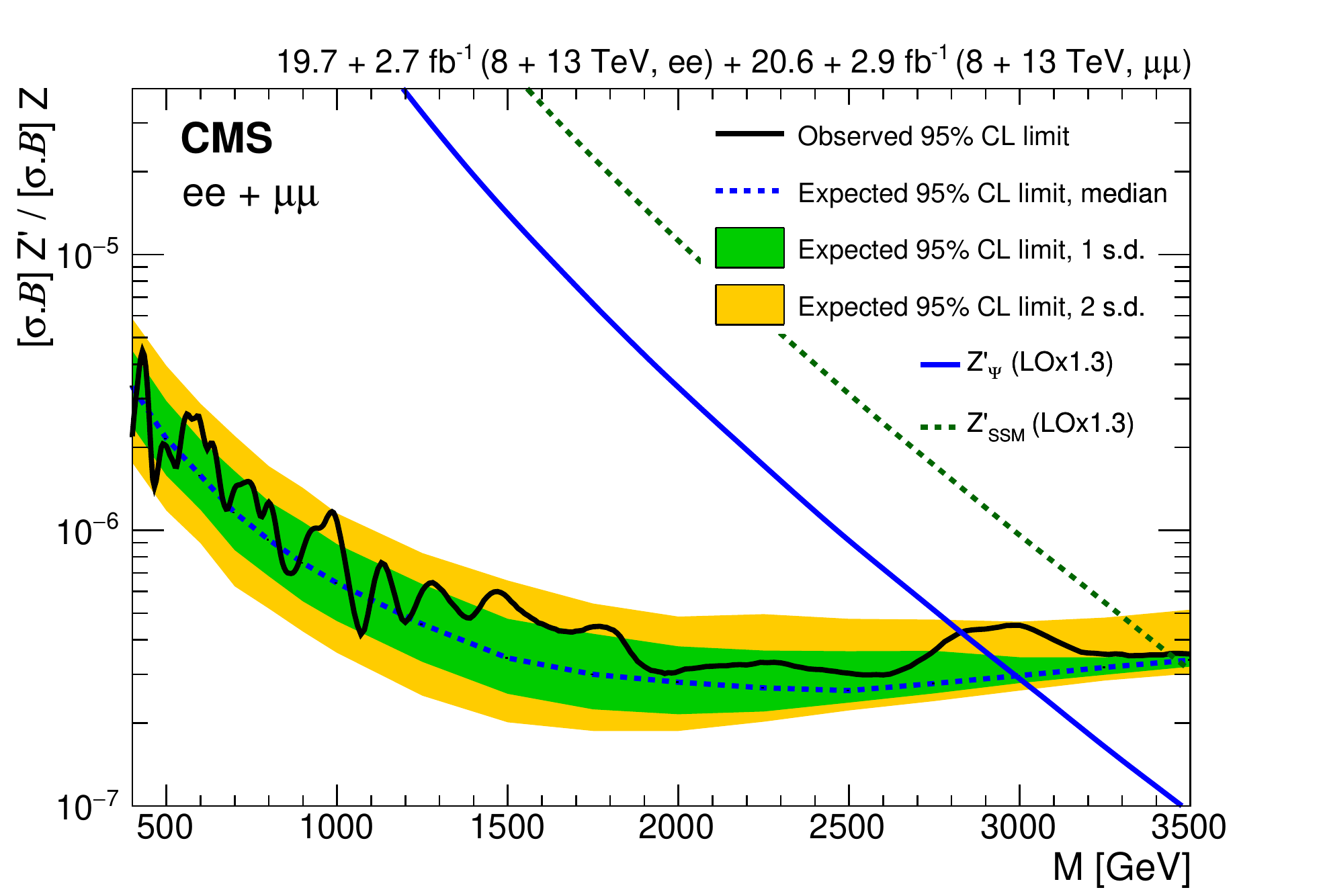}
\includegraphics[width=0.49\textwidth]{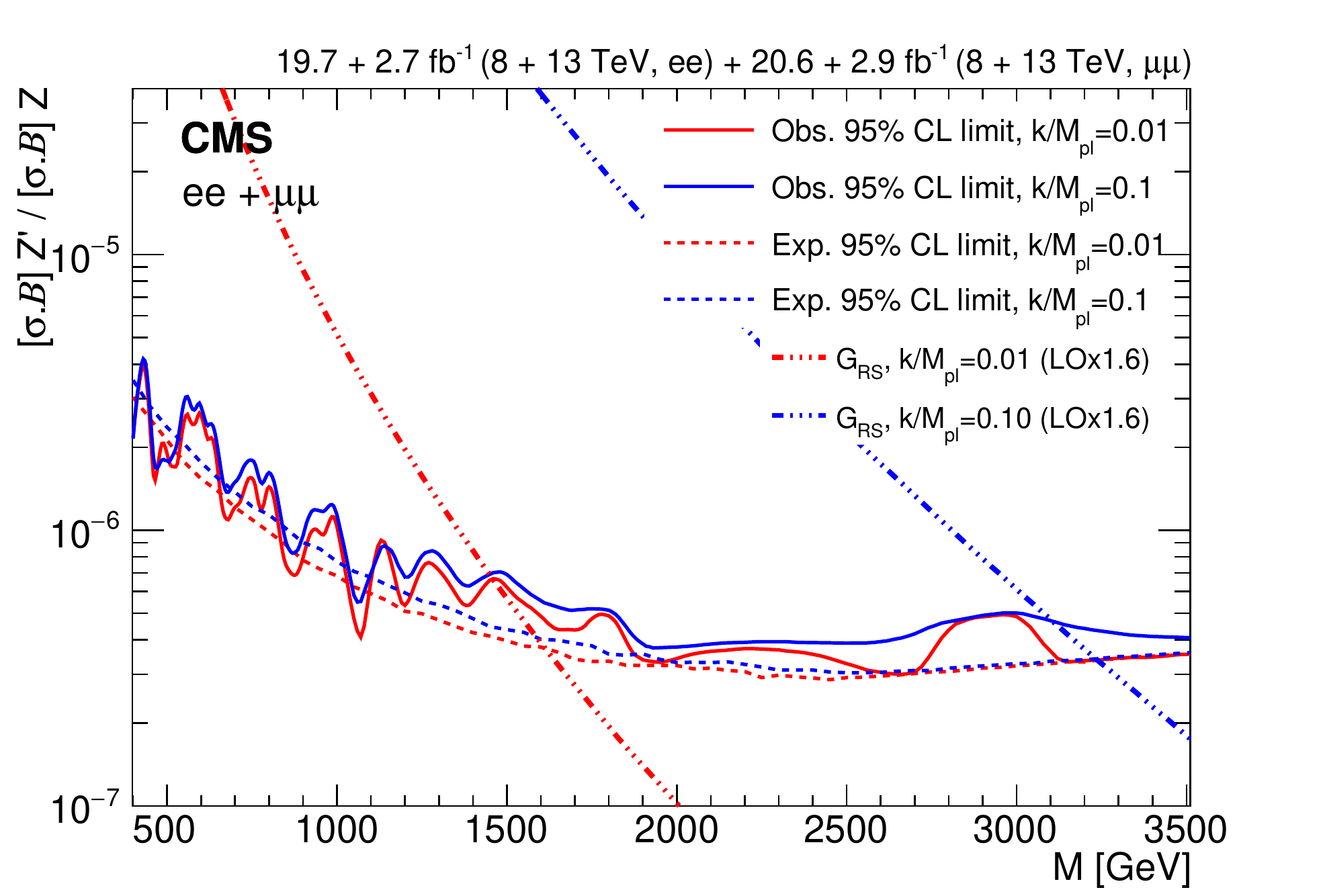}
 \caption{The 95\% CL upper limits on the product of production cross section
   and branching fraction for (\cmsLeft) a spin-1 resonance with a width equal
   to 0.6\% of the resonance mass and (\cmsRight) for a spin-2 RS graviton,
   both relative to the product of production cross section and
   branching fraction for a Z boson, for the combined dielectron and
   dimuon channels and combined 8 and 13\TeV data.  For the spin-1
   results (\cmsLeft plot), the shaded bands correspond to the 68 and 95\%
   quantiles for the expected limits, and theoretical predictions are
   shown for the spin-1 \ZPSSM and \ZPPSI resonances.  For the
   spin-2 results (\cmsRight plot), observed limits, expected limits, and
   theoretical predictions are shown for values of the coupling
   parameter $k/\overline{M}_\mathrm{Pl}=0.01$ and 0.10.  }
\label{fig:limits6}
\end{figure}

The 95\% CL upper limits on $R_{\sigma}$ for the combined 8 and 13\TeV
data are shown in Fig.~\ref{fig:limits5} for the individual dielectron
and dimuon channels, and in Fig.~\ref{fig:limits6} (\cmsLeft) for the
combination of the two channels.  Figure~\ref{fig:limits6} (\cmsRight)
shows the 95\% CL upper limits on the product of production cross section and branching
fraction for an RS graviton, normalized to the same quantity for the Z
boson, for the combination of the 8 and 13\TeV data and of the two
dilepton channels.

The 95\% CL lower limits on the masses of the \ZPSSM and \ZPPSI
bosons are presented in Table~\ref{tab:massLimitsSpin1}, along with
the expected results. Table~\ref{tab:massLimitsSpin2} presents the
corresponding limits for an RS graviton with coupling parameters 0.01
and 0.10. In each case the limit appropriate to the width of the boson
is used. For example the \ZPSSM boson mass limits are calculated
using a width of 3\%.  The cross section as a function of mass is
calculated at LO using the \PYTHIA~8.2 program with the NNPDF2.3 PDFs.
As the limits in this Letter are obtained on the on-shell cross
section and the \PYTHIA\ event generator includes off-shell effects,
the cross section is calculated in a mass window of
$\pm5\%$~$\sqrt{s}$ centred on the resonance mass, following the
advice of Ref.~\cite{Accomando:2013sfa}. The validity of this
procedure for the $\ZPSSM$ and $\ZPPSI$ bosons was explicitly checked
in Ref.~\cite{Accomando:2013sfa} and was found to be accurate at the 5--7\% level. To account for NLO effects,
the cross sections are multiplied by a $K$-factor of 1.3 for $\cPZpr$
models and 1.6 for RS graviton models~\cite{Mathews:2005bw}, with the
$K$-factor for $\cPZpr$ models obtained by comparing \POWHEG and
\PYTHIA cross sections for SM Drell--Yan production. These same
comments apply for the theoretical predictions shown in
Figs.~\ref{fig:limits2}--\ref{fig:limits6}.  For the \ZPSSM and
\ZPPSI bosons, we obtain lower mass limits of $\limitZssm$ and
$\limitZpsi\TeV$, respectively. The lower mass limit obtained for the
RS graviton is $\limitGone$ ($\limitGthree$)\TeV for a coupling
parameter of 0.01 (0.10).

\begin{table*}[!hbt]
\centering
\topcaption{The observed and expected 95\% CL lower limits on the masses
  of spin-1 \ZPSSM and \ZPPSI bosons for the combination of the 8
  and 13\TeV data, assuming a signal width of 0.6\% of the resonance
  mass for \ZPPSI and 3\% for \ZPSSM. }
\begin{tabular}{ccc{c}@{\hspace*{5pt}}cc}
\multirow{2}{*}{Channel}  & \multicolumn{2}{c}{\ZPSSM} && \multicolumn{2}{c}{\ZPPSI}  \\\cline{2-3}\cline{5-6}
                          & Obs. (\TeVns) & Exp. (\TeVns)      && Obs. (\TeVns)  & Exp. (\TeVns)      \\\hline
$\Pe\Pe$                        & 2.95   & 3.11    && 2.60        & 2.67            \\
$\mu^+\mu^-$              & 3.22   & 3.23    && 2.77        & 2.77            \\
$\Pe\Pe$ + $\mu^+\mu^-$         & 3.37   & 3.45    && 2.82        & 2.98            \\
$\Pe\Pe$ + $\mu^+\mu^-$  13\TeV only     &    3.18   &    3.35   &&   2.70  &  2.82        \\
\end{tabular}
\label{tab:massLimitsSpin1}

\end{table*}

\begin{table*}[!hbt]
\centering
\topcaption{The observed and expected 95\% CL lower limits on the masses
  of spin-2 Kaluza--Klein gravitons in the Randall--Sundrum model for the combination of the 8 and 13\TeV data, assuming
  two values of the coupling parameter, $k/\overline{M}_\mathrm{Pl}$.}
\begin{tabular}{ccc{c}@{\hspace*{5pt}}cc}
\multirow{2}{*}{Channel}  & \multicolumn{2}{c}{$\GKK$ ($k/\overline{M}_\mathrm{Pl}=0.01$)}&& \multicolumn{2}{c}{$\GKK$ ($k/\overline{M}_\mathrm{Pl}=0.10$)}    \\\cline{2-3}\cline{5-6}
                          & Obs. (\TeVns)  & Exp. (\TeVns)         &&
                          Obs. (\TeVns)  & Exp. (\TeVns)
                          \\\hline
$\Pe\Pe$  & 1.46 & 1.48 && 2.78  & 2.93 \\
$\mu^+\mu^-$    & 1.26 & 1.41 && 3.03 & 3.03 \\
$\Pe\Pe$ + $\mu^+\mu^-$   &  1.46     &  1.61     &&    3.11    &      3.23      \\
$\Pe\Pe$ + $\mu^+\mu^-$  13\TeV only       & 1.38        & 1.45       && 2.98        & 3.15                   \\
\end{tabular}
\label{tab:massLimitsSpin2}

\end{table*}

\section{Summary}

A search for narrow resonances in dielectron and dimuon invariant mass
spectra has been performed using data obtained from proton-proton
collisions at $\sqrt{s}=13\TeV$. The integrated luminosity for the
dielectron sample is $\anaLumiee\fbinv$ and for the dimuon sample
$\anaLumimumu\fbinv$. The sensitivity of the search is increased by
combining these data with a previously analysed set of data obtained
at $\sqrt{s}=8\TeV$ and corresponding to a luminosity of 20\fbinv. No
evidence for non-standard-model physics is found, either in the 13\TeV
data set alone, or in the combined data set.  Upper limits at 95\%
confidence level on the product of production cross section and
branching fraction have also been calculated in a model-independent
manner to enable interpretation in models predicting a narrow
dielectron or dimuon resonance structure.

Limits are set on the masses of hypothetical particles that could
appear in new-physics scenarios.  For the \ZPSSM particle, which
arises in the sequential standard model, and for the superstring
inspired \ZPPSI particle, 95\% confidence level lower mass limits for
the combined data sets and combined channels are found to be
$\limitZssm$ and $\limitZpsi\TeV$, respectively. The corresponding
limits for Kaluza--Klein gravitons arising in the Randall--Sundrum
model of extra dimensions with coupling parameters 0.01 and 0.10 are
$\limitGone$ and $\limitGthree\TeV$, respectively. These results
significantly exceed the limits based on the 8\TeV LHC data.

\begin{acknowledgments}
We congratulate our colleagues in the CERN accelerator departments for the excellent performance of the LHC and thank the technical and administrative staffs at CERN and at other CMS institutes for their contributions to the success of the CMS effort. In addition, we gratefully acknowledge the computing centres and personnel of the Worldwide LHC Computing Grid for delivering so effectively the computing infrastructure essential to our analyses. Finally, we acknowledge the enduring support for the construction and operation of the LHC and the CMS detector provided by the following funding agencies: BMWFW and FWF (Austria); FNRS and FWO (Belgium); CNPq, CAPES, FAPERJ, and FAPESP (Brazil); MES (Bulgaria); CERN; CAS, MoST, and NSFC (China); COLCIENCIAS (Colombia); MSES and CSF (Croatia); RPF (Cyprus); SENESCYT (Ecuador); MoER, ERC IUT and ERDF (Estonia); Academy of Finland, MEC, and HIP (Finland); CEA and CNRS/IN2P3 (France); BMBF, DFG, and HGF (Germany); GSRT (Greece); OTKA and NIH (Hungary); DAE and DST (India); IPM (Iran); SFI (Ireland); INFN (Italy); MSIP and NRF (Republic of Korea); LAS (Lithuania); MOE and UM (Malaysia); BUAP, CINVESTAV, CONACYT, LNS, SEP, and UASLP-FAI (Mexico); MBIE (New Zealand); PAEC (Pakistan); MSHE and NSC (Poland); FCT (Portugal); JINR (Dubna); MON, RosAtom, RAS and RFBR (Russia); MESTD (Serbia); SEIDI and CPAN (Spain); Swiss Funding Agencies (Switzerland); MST (Taipei); ThEPCenter, IPST, STAR and NSTDA (Thailand); TUBITAK and TAEK (Turkey); NASU and SFFR (Ukraine); STFC (United Kingdom); DOE and NSF (USA).

Individuals have received support from the Marie-Curie programme and the European Research Council and EPLANET (European Union); the Leventis Foundation; the A. P. Sloan Foundation; the Alexander von Humboldt Foundation; the Belgian Federal Science Policy Office; the Fonds pour la Formation \`a la Recherche dans l'Industrie et dans l'Agriculture (FRIA-Belgium); the Agentschap voor Innovatie door Wetenschap en Technologie (IWT-Belgium); the Ministry of Education, Youth and Sports (MEYS) of the Czech Republic; the Council of Science and Industrial Research, India; the HOMING PLUS programme of the Foundation for Polish Science, cofinanced from European Union, Regional Development Fund, the Mobility Plus programme of the Ministry of Science and Higher Education, the National Science Center (Poland), contracts Harmonia 2014/14/M/ST2/00428, Opus 2013/11/B/ST2/04202, 2014/13/B/ST2/02543 and 2014/15/B/ST2/03998, Sonata-bis 2012/07/E/ST2/01406; the Thalis and Aristeia programmes cofinanced by EU-ESF and the Greek NSRF; the National Priorities Research Program by Qatar National Research Fund; the Programa Clar\'in-COFUND del Principado de Asturias; the Rachadapisek Sompot Fund for Postdoctoral Fellowship, Chulalongkorn University and the Chulalongkorn Academic into Its 2nd Century Project Advancement Project (Thailand); and the Welch Foundation, contract C-1845.
\end{acknowledgments}
\bibliography{auto_generated}
\cleardoublepage \appendix\section{The CMS Collaboration \label{app:collab}}\begin{sloppypar}\hyphenpenalty=5000\widowpenalty=500\clubpenalty=5000\input{EXO-15-005-authorlist.tex}\end{sloppypar}
\end{document}

%% file: EXO-15-005-authorlist.tex
\textbf{Yerevan Physics Institute,  Yerevan,  Armenia}\\*[0pt]
V.~Khachatryan, A.M.~Sirunyan, A.~Tumasyan
\vskip\cmsinstskip
\textbf{Institut f\"{u}r Hochenergiephysik der OeAW,  Wien,  Austria}\\*[0pt]
W.~Adam, E.~Asilar, T.~Bergauer, J.~Brandstetter, E.~Brondolin, M.~Dragicevic, J.~Er\"{o}, M.~Flechl, M.~Friedl, R.~Fr\"{u}hwirth\cmsAuthorMark{1}, V.M.~Ghete, C.~Hartl, N.~H\"{o}rmann, J.~Hrubec, M.~Jeitler\cmsAuthorMark{1}, A.~K\"{o}nig, I.~Kr\"{a}tschmer, D.~Liko, T.~Matsushita, I.~Mikulec, D.~Rabady, N.~Rad, B.~Rahbaran, H.~Rohringer, J.~Schieck\cmsAuthorMark{1}, J.~Strauss, W.~Treberer-Treberspurg, W.~Waltenberger, C.-E.~Wulz\cmsAuthorMark{1}
% ...
\vskip\cmsinstskip
% ...
1:~~Also at Vienna University of Technology, Vienna, Austria\\
% ...
